\documentstyle[pre,aps,multicol,epsfig,epsf]{revtex}
\renewcommand{\narrowtext}{\begin{multicols}{2} \global\columnwidth20.5pc}
\renewcommand{\widetext}{\end{multicols} \global\columnwidth42.5pc}

\def\Xop{\hat{X}}
\def\Top{\hat{T}}
\def\i{|i\rangle}
\def\sbar{|\bar{s}\rangle}
\begin{document}
\draft
\title{\Large\bf {Fast Quantum Search Algorithms in Protein Sequence
Comparison - Quantum Biocomputing}}

\author{\Large Lloyd C.L. Hollenberg}

\address{Max--Planck--Institut f\"ur Kernphysik, Saupfercheckweg 1,
Heidelberg D-69117, GERMANY.}
\date{24 February, 2000}
\maketitle

\begin{abstract}
Quantum search algorithms are considered in the context of protein
sequence comparison in biocomputing. Given a sample protein
sequence of length $m$ (i.e $m$ residues), the problem considered
is to find an optimal match in a large database containing $N$ residues.
Initially, Grover's quantum search algorithm is applied to a
simple illustrative case - namely where the database forms a
complete set of states over the $2^m$ basis states of a $m$ qubit
register, and thus is known to contain the exact sequence of
interest. This example demonstrates explicitly the typical
$O(\sqrt{N})$ speedup on the classical $O(N)$ requirements. An
algorithm is then presented for the (more realistic) case where
the database may contain repeat sequences, and may not necessarily
contain an exact match to the sample sequence. In terms of
minimizing the Hamming distance between the sample sequence and
the database subsequences the algorithm finds an optimal
alignment, in $O(\sqrt{N})$ steps, by employing an extension of
Grover's algorithm, due to Boyer, Brassard, H$\o$yer and Tapp for
the case when the number of matches is not a priori known.
\end{abstract}

\pacs{PACS number(s): 03.67.Lx,03.67.-a,87.15.Cc}

\narrowtext

The fantastic possibilities of quantum parallelism in computing,
suggested by the convergence of quantum mechanics and information
theory in the past two decades, are fast being enumerated in the
guise of quantum algorithms. First, and foremost, among these is
the factoring algorithm of Shor\cite{shor}, which provided great
impetus to the field of quantum computing. Shor's algorithm
applied to a given number $N$ requires $O((\log N)^3)$ steps, and
represents an exponential speed-up over the best classical
algorithms. Another important result, due to Grover\cite{grover},
was the discovery of a quantum search algorithm for finding a
particular element in an unordered set of $N$ elements in only
$O(\sqrt{N})$ steps - a significant improvement over the classical
cost $O(N)$.

In this paper the application of quantum search algorithms to an
important problem at the heart of biocomputing (or
bioinformatics), that of protein sequence comparison and
alignment, is considered. As the mapping and sequencing of the
human genome (some $3\times 10^9$ base pairs) nears completion,
the relatively new field of biocomputing has become obvious in its
importance to the quantitative analysis of this vast amount of
data. Some fundamental tasks in biocomputing involving sequence
analysis include: searching databases in order to compare a new
sub-sequence to existing sequences, inferring protein sequence
from DNA sequence, and calculation of sequence alignment in the
analysis of protein structure and function. A tremendous amount of
computing is required, much of which is devoted to search-type
problems, either directly in large databases, or in configuration
space of alignment possibilities. While it is possible that all of
these problems may be amenable to quantum algorithmic speed-up, it
is explicitly demonstrated in this work how the fundamental task
of sequence alignment can be approached using a quantum computer.
Indeed, this problem is a very natural application of the quantum
search algorithm (perhaps a strange reflection of the possibility
that the machinery of DNA itself may actually function using
quantum search algorithms\cite{patel}).

In general terms Grover's search algorithm relies on the existence
of a quantum computer $Q$ operating using an oracle function, F.
The set of search possibilities is represented by states in the
Hilbert space of $Q$. The oracle function simply tests whether a
given state is the actual target state. Grover found a unitary
operator $U$ (involving the oracle function test) which evolves
the quantum computer in such a way that the amplitude of the
target state in the wave function of $Q$ is amplified.
Furthermore, Grover showed that there exists a number $k <
\sqrt{N}$, such that after $k$ applications of $U$, the
probability of finding the target state is at least 1/2.
Subsequently, Boyer, Brassard, H$\o$yer and Tapp (BBHT) proved a
tighter bound: one must iterate the algorithm on average at least
$\left(\sin {\pi\over 8}\right)\sqrt{N}$ times to achieve a
probability of 1/2 for finding the target\cite{BBHT}.

To begin the application of quantum search algorithms to protein
sequence analysis, the problem of sequence alignment to a large
database of sequence domains is considered. That is, given a
sample sequence the task is to find out the location in the
database of an exact or closest match (with respect to some
defined measure). Application of the Grover algorithm directly to
this search task would cause trouble immediately because, by
definition, it is not known if the target exists in the database,
or if it actually exists multiple times. If there are actually
$N_t$ solutions, the number of iterations required to find a
solution with probability 1/2 is $\left(\sin {\pi\over
8}\right)\sqrt{N/N_t}$\cite{BBHT}. Thus, if one does not know the
number of solutions at the outset, the computer may inadvertently
be halted when the amplitude of the target states is very small.
This happens because the process of amplitude amplification is not
monotonic, but rather oscillates with the number of iterations.
Fortunately, this difficult impasse has been solved by BBHT and
they provide an algorithm, based on Grover's algorithm as a
subroutine, for finding a solution in the case where the number of
solutions is unknown\cite{BBHT}. This result allows for the
application of quantum search algorithms to the field of
biocomputing.

In terms of protein sequences, the human genome is composed of
about 150,000 domains, each containing on average 300 residues
(amino acids). An interesting feature of approaching the sequence
analysis problem using a quantum computer is that the entire
database could in principle be stored in a single wave function
superposition, and then be presented simultaneously for
inspection. To illustrate the basic idea, a very simple case of
sequence comparison is initially considered, followed by a more
realistic problem later. Consider a database, $D$, constructed
from the domains of the human genome placed end-to-end, so that a
continuous list of $N$ residues, $D=\{R_0,R_1,...,R_{N-1}\}$, is
created. Independently, a sample sequence is given $s =
\{r_0,r_1,...,r_{m-1}\}$ composed of $m$ residues; the task is to
compare this with the database. Each residue is labeled by a
letter of the 20-letter amino acid alphabet, so in order to encode
the database 5 bits per residue is needed. Thus, the residues
$R_i$ and $r_i$ are represented by bit strings,
$\prod_{\alpha=0}^{4}\,B_{i\alpha}$ and
$\prod_{\alpha=0}^{4}\,b_{i\alpha}$, respectively.

The quantum computer to analyze this system is composed of two
registers, with number of qubits $Q_1$ and $Q_2$, respectively.
The bit-wise representation of the protein sequences will be
encoded into the qubits of this system. Leaving issues of data
transfer aside, the entire database is represented by a quantum
superposition over the two registers:
\begin{equation}
|\Psi_D\rangle \equiv {1\over \sqrt {N-m+1}
}\,\sum_{i=0}^{N-m}\,|\phi_i\rangle\,\otimes|i\rangle,
\end{equation}
where all the consecutive sub-sequences in the database of length
$m$ are encoded in the first register with $Q_1=5 m$ as
\begin{equation}
|\phi_i\rangle =
\prod_{\alpha=i}^{i+m-1}\,\prod_{\beta=0}^{4}|B_{\alpha\beta}\rangle\equiv
\prod_{\alpha=0}^{5m-1}|q_{i\alpha}\rangle.
\end{equation}
That is, from from the database of length $N$ residues, $N-m+1$
sub-sequences of length $m$ are constructed by moving along from
the first position (allowing domain crossing). Position
information of the sub-sequences is meanwhile tagged explicitly by
binary numbers, $|i\rangle$, in the second register, and is
accessed by an operator, $\Xop$, acting in the Hilbert space of
the second register, which gives the position as $\Xop \,|i\rangle
= i\,|i\rangle$ ($0\le i\le N-m$). In order that this register can
encode all positions $Q_2$ must satisfy $2^{Q_2}> N-m$. The number
of qubits required in this register is relatively small: taking
the database size to be that for the number of residues in the
human genome implies $Q_2=26$ suffices. In the first register,
typical sequence comparison problems require $m\sim O(300)$.

The next step in the initialization process is the coding of a
table, $T[0...N-m]$, into the quantum state, which measures the
difference between the database states $|\phi_i\rangle$ and the
sample sequence state in terms on the total number of bit flips
required to transform any database state into the sample sequence.
In other words, $T[0...N-m]$ is the set of Hamming distances.
Remarkably, the set of Hamming distances for the {\em entire}
database can be created by simply acting on each qubit of the
computer with a CNOT operation with respect to the sample sequence
state:
\begin{equation}
|\Psi_H\rangle = U_{\rm CNOT}(s)\,|\Psi_D\rangle \equiv {1\over
\sqrt {N-m+1}
}\,\sum_{i=0}^{N-m}\,|\bar{\phi_i}\rangle\,\otimes|i\rangle.
\end{equation}
Denoting the individual qubits of the ``Hamming states''
$|\bar{\phi_i}\rangle$ by
\begin{equation}
|\bar{\phi_i}\rangle =
\prod_{\alpha=0}^{5m-1}|\bar{q}_{i\alpha}\rangle,
\end{equation}
an operator, $\Top$, is introduced which, acting on a state
$|\bar{\phi_i}\rangle$, gives the Hamming distance table value
$T[i]$ as:
\begin{equation}
\Top \,:\,\Top\,|\bar{\phi_i}\rangle = T[i]\,|\bar{\phi_i}\rangle,
\quad T[i] =\sum_{\alpha=0}^{5m-1}\bar{q}_{i\alpha}.
\end{equation}

With the computer design completed and initialized, a simple
search problem can be defined in order to demonstrate how the
computer works. First, the database is taken to be of length
$N=2^m+m-1$ so that there are exactly $2^m$ states in the
superposition, and furthermore demand that all these states are
distinct. The problem is to search the database for the
sub-sequence $s$, which occurs exactly once, but at an unknown
location. Classically, this would require $O(N)$ steps. However,
by using Grover's search algorithm, the match can be found in
$O(\sqrt{N})$ steps. In this example, the database decomposition
has been artificially arranged to be over a complete set of states
of the first register, which means that Grover's search algorithm
can be applied directly.

The problem defined by Grover\cite{grover} has been modified
slightly, but the applicability of the search algorithm remains.
The original problem was defined in terms of an oracle function,
$F(x)$, over a set of values $x\in\{0,\dots,N-1\}$, which is zero
everywhere except at some value $t$, the target of the search,
where $F(t)=1$. The sequence comparison problem here has been
re-structured so that a value of $x$ represents a subsequence of
the database, and the oracle function is just a direct comparison
with the sample sequence. In a sense, the black box nature of the
oracle function has been simplified, at the cost of increasing the
complexity of the initial wave function with position information.
It remains to be seen whether this is a feasible way of coding a
sequence database. Of course, an alternative is to sweep all
details of the database look-up and comparison into the oracle
function. The difference is subtle, and perhaps non-trivial in
practice. The advantage of the latter approach might be in the
initialization of the quantum computer state. The algorithms
presented here would still apply in this case.

In the computer design defined here, Grover's search algorithm is
applied to the first register containing the sub-sequence state
superposition. The problem is to find the state $\sbar = U_{\rm
CNOT}|s\rangle = |0\dots0\rangle$ (zeros in all $m$ qubits of the
first register) with table value $T[i_s]=0$, occurring at position
$i_s$ (as yet unknown). Once the state is found, the location of
the sequence in the database can be determined by making a
measurement of $\Xop$ on the second register.

To illustrate the working of the algorithm the geometrical
picture\cite{grover2,farhi,chen}, which is particularly
transparent, is applied to this framework. The search algorithm is
initiated by decomposing the state $|\Psi_H\rangle$ into
orthogonal components with respect to $\sbar$ as
\begin{equation}
|\Psi_H\rangle = \sqrt{N-m\over N-m+1}\,|R\rangle + {1\over
\sqrt{N-m+1}}\,|S\rangle,
\end{equation}
where
\begin{eqnarray}
|S\rangle  &=& \sbar\,\otimes|i_s\rangle\nonumber\\
|R\rangle &=& {1\over \sqrt{N-m}}\,\sum_{i\neq i_s}
|\bar{\phi_i}\rangle\otimes \i .
\end{eqnarray}

The evolution of the quantum computer representing the search
algorithm occurs in the first register, the second register lying
dormant, yet through quantum entanglement carrying the position
information required at the end. The operator $U$ is constructed
from reflection operators in the Hilbert space of the first
register,
\begin{eqnarray}
I_S &=& 1 - 2\,|S\rangle\langle S|\nonumber\\
I_H &=& 1-2\, |\Psi_H\rangle\langle\Psi_H|.
\end{eqnarray}
The operator $I_S$ contains the query to the oracle function,
$F(i)$, and acts on the Hamming states $|\bar{\phi_i}\rangle$ with
a phase shift dependent on the search criteria $T[i_s] = 0$:
\begin{equation}
I_S\,|\bar{\phi_i}\rangle = (-1)^{F(i)}\,|\bar{\phi_i}\rangle =
\left\{
\begin{array}{lll}
-\,|\bar{\phi_i}\rangle & & {\rm if\,\,} T[i]=0\\& &\\
|\bar{\phi_i}\rangle    & & {\rm otherwise}.
\end{array} \right.
\end{equation}

In terms of these reflection operators, the unitary operator
evolving the system through one step of the search algorithm is
given by $U = - I_H\,I_S$. The evolution of the computer proceeds
through application of the operator, $U$, a number of times on the
initial state, $|\Psi_H\rangle$. The effect of this evolution is
to amplify the component of the target state, $|S\rangle$ in the
superposition. It is important to understand the nature of this
process in order to appreciate how the quantum computer functions.
To see this point it is convenient to express $U$ in the
representation of the subspace $\{|S\rangle,|R\rangle\}$:
\begin{eqnarray}
U = \left[
\begin{array}{ccc}
{N-m-1\over N-m+1}& & {2\sqrt{N-m} \over N-m+1}\\ & &\\
-{2\sqrt{N-m} \over N-m+1} & & {N-m-1\over N-m+1}
\end{array}
\right]
= \left[
\begin{array}{ccc}
\cos\theta & & \sin\theta \\ & &\\ -\sin\theta & & \cos\theta
\end{array}
\right] \label{Umatrix}.
\end{eqnarray}
where $\sin\theta \equiv 2\sqrt{N-m}/ (N-m+1)$.

After $k$ steps of the algorithm the state of the computer is
given by
\begin{equation}
|\Psi_k\rangle = U^k\,|\Psi_H\rangle =
\sum_{i=0}^{N-m}\,c_i^{(k)}\,|\bar{\phi_i}\rangle\otimes\i.
\end{equation}
The amplitude of the target state, $c_{i_s}^{(k)}$, can be easily
calculated using the matrix representation for $U$. One obtains:
\begin{equation}
c_{i_s}^{(k)} = \cos\left(k\,\theta -
\alpha\right),\quad\quad\cos\alpha \equiv {1\over \sqrt{N-m+1}}.
\end{equation}

The component along $|S\rangle$ is amplified to near unity at
$k_{\rm max}\sim {\pi\over 4}\sqrt{N}$ (for $N>>m$). A measurement
of $\Top$ on the first register will give a result $T[i]$ with
probability $|c_i^{(k)}|^2$. If $T[i]=0$ then the algorithm has
succeeded - i.e the sample sequence has been found - and a
subsequent measurement of $\Xop$ in the second register will give
the position, $i_s$, of the sequence in the database. A crucial
point is that one has to be careful interpreting the number of
steps required to obtain a successful outcome - merely increasing
the number of steps beyond $k_{\rm max}$ does not improve the
chances of success because the amplification is not monotonic.
Indeed, the probability of success actually decreases when $k_{\rm
max}$ is exceeded. The search may therefore have to be run several
times, however, for large $N$ the savings in computer time
compared to a classical computer are clear, even if the search is
repeated several times.

While the above example serves to display the potential of quantum
search algorithms in the context of sequence matching to a large
database, it does not contain an important concept in
bioinformatics - optimal alignment. Generally, the sample sequence
may not be contained exactly in the database, and so one is
interested in how close is the best match (or matches) with
respect to a well defined distance measure. Often this measure
involves editing of strings by insertion of gaps in order to
minimize the distance; in practice this process is very
complicated. In the first instance, the problem is extended to
that of finding an optimal alignment with respect to the Hamming
distance, without editing of sequences (which can be incorporated
at a later stage).

Let us first define the problem using, as far as possible, the
same notation as previously. The database is taken to be of size
$N>>m$, but the restriction that the set of database sub-sequence
states is equal to $2^m$ is relaxed, and the possibility is
allowed that the set of sub-sequences may contain repeats, and,
more importantly, may or may not contain the sample sequence. The
problem then is to find an optimal alignment of the sample
sequence to a sub-sequence in the database. An optimal alignment
here is defined in the sense of finding the smallest Hamming
distance $T[i]$ with respect to the sample sequence state.

In terms of our quantum computer, the database state in this case
is also described by the state $|\Psi_D\rangle$. An important
point is that the state is still normalised by the factor $1/\sqrt
{N-m+1}$ because the repeats occur at different locations, and
thus each state in the product space of the two registers is
distinct. The introduction of the position register $Q_2$ has
ensured this. Using the CNOT operation on $|\Psi_D\rangle$ the
superposition, $|\Psi_H\rangle$, of Hamming states is once again
obtained. The algorithm strategy is to search for alignments of
increasing Hamming distance. At the start of each search it is not
known how many solutions exist, or if there exist matches at all,
and so Grover's algorithm cannot be used directly. However, we use
now the extension of Grover's algorithm due to Boyer, Brassard,
H$\o$yer and Tapp, which performs a search with an a priori
unknown number of solutions $N_t$, and finds a match (if it
exists) in $O(\sqrt{N/N_t})$ steps\cite{BBHT}. During the course
of the algorithm the computer's evolution must be tailored to
accommodate the fact that the search is now based on all the
target states that satisfy $T[i] = n$ where $n$ is some
pre-defined Hamming distance determined by the algorithm. In order
to apply the search algorithm in this case the operator
$I_S\rightarrow I_S(n)$ is modified such that:
\begin{equation}
I_S(n)\,|\bar{\phi_i}\rangle = \left\{
\begin{array}{lll}
-\,|\bar{\phi_i}\rangle & & {\rm if\,\,} T[i]=n\\& &\\
|\bar{\phi_i}\rangle    & & {\rm otherwise}.
\end{array} \right.
\end{equation}

At each iteration the BBHT algorithm is employed, with a repeat
index $r$ as a pre-determined measure of the search confidence
level.

The optimal alignment algorithm is as follows:

\begin{enumerate}
\item 0th iteration: search for an occurrence of the state with
zero Hamming distance, $T[i] = 0$. If successful measure position
and exit, if unsuccessful after $r$ repeats of the BBHT search
algorithm go to the next iteration.
\item $n$th iteration: search for a state with $T[i]=n$ using
$U=-I_H\,I_S(n)$. If successful locate position and exit, if
unsuccessful after $r$ repeats of the BBHT search algorithm go to
the next iteration, by setting $n\rightarrow n+1$.
\item Upon exit at some iteration $n=k$, one optimal alignment
$T[i_k]=k$, and its position $i_k$ has been found.
\end{enumerate}

The total number of steps required is $O(r k \sqrt{N})$,
discounting the effect of sequence repeats (which reduces the
required number of iterations). At more cost a sub-loop may be
introduced to search for the other optimally aligned sequences. In
practice, the number of iterations required is $k<<m$, as one
would determine a maximum Hamming distance on biological grounds,
beyond which searching for an aligned state is pointless.

While the focus of this paper has been on protein sequence
comparison, the framework can be easily translated into that for
nucleotide sequence comparison in DNA. In this case representing
the four letter nucleotide alphabet requires only two qubits.

Although only the algorithmic aspect of the application of quantum
computing to sequence analysis has been dealt with here, an
obvious point to raise is the feasibility of building such a
device. With the ever increasing ability to manipulate systems at
the quantum level there has been great progress in the
demonstration of quantum computation at the two qubit level.
Quantum logic gates were demonstrated using ion
traps\cite{iongates} in 1995, and two years later in nuclear
magnetic resonance (NMR) systems\cite{NMRgates}. In 1998 the
actual experimental realization of a quantum computer solving
Deutsch's problem was reported by two groups using
NMR\cite{jones,chuang}. This was closely followed by NMR
implementations of the quantum search
algorithm\cite{chuangsearch,jonessearch}. Of course, a realistic
quantum computer needs to be scaled up significantly on these two
qubit configurations. Perhaps the most promising prospect for a
scalable quantum computer capable of running the algorithms
presented here is based on the solid state design of
Kane\cite{kane}. The creation of a superposition representing the
human genome database would be another considerable challenge.

To conclude, in this work the application of quantum search
algorithms in the context of biocomputing has been studied, at
least at the rather simple level of sequence alignment with
respect to the Hamming distance. Actual alignment problems would
include alignment through editing of sequences - i.e. insertion of
gaps. It is quite possible that this procedure can be achieved
using a multi-qubit representation (which includes gap characters)
within the quantum search algorithm process by suitable choice of
qubit evolution operators. Work in this direction is in progress.

The author wishes to thank S. O'Donoghue for many helpful
discussions. Thanks also to G. Akemann and L. Benet Fernandez for
kindly reading through the manuscript, and the theory group at
MPIK for their hospitality. This work was supported by the
Alexander von Humboldt Foundation.

\widetext
\end{document}